# A Survey of Credit Card Fraud Detection Techniques: Data and Technique Oriented Perspective

SamanehSorournejad[1], Zahra Zojaji[2], Reza Ebrahimi Atani[3], Amir Hassan Monadjemi[4]

[1]Department of Information Technology, University of Guilan, Iran
sorournejad@yahoo.com
[2]Department of Computer Engineering, Amirkabir University of Technology
z.zojaji@aut.ac.ir
[3]Department of Computer Engineering, University of Guilan, Rasht, Iran
rebrahimi@guilan.ac.ir
[4]Department of Computer Engineering, University of Isfahan, Isfahan, Iran
monadjemi@eng.ui.ac.ir

**Abstract**

Credit card plays a very important rule in today's economy. It becomes an unavoidable part of household, business and global activities. Although using credit cards provides enormous benefits when used carefully and responsibly,significant credit and financial damagesmay be causedby fraudulent activities. Many techniques have been proposed to confront thegrowthin credit card fraud. However, all of these techniques have the same goal of avoiding the credit card fraud; each one has its own drawbacks, advantages and characteristics. In this paper, after investigating difficultiesof credit card fraud detection, we seek to review the state of the art in credit card fraud detection techniques, datasets and evaluation criteria.The advantages and disadvantages of fraud detection methods are enumerated and compared.Furthermore, a classification of mentioned techniques into two main fraud detection approaches, namely, misuses (supervised) and anomaly detection (unsupervised) is presented. Again, a classification of techniques is proposed based on capability to process the numerical and categorical datasets. Different datasets used in literatureare then described and grouped into real and synthesized data and the effective and common attributesare extracted for further usage.Moreover, evaluation employed criterions in literature are collected and discussed.Consequently, open issues for credit card fraud detection are explained as guidelinesfor new researchers.

## 1. Introduction
At the current state of the world, financial organizations expand the availability of financial facilitiesbyemployingof innovative servicessuch ascredit cards, Automated Teller Machines (ATM), internet and mobile banking services. Besides, along with the rapid advances of e-commerce,the use of credit card has become a convenience and necessary part of financial life. Credit card is a payment cardsupplied to customers as a system of payment. There are lots of advantages in usingcredit cards such as:

- **Ease of purchase**
  Credit cards can make life easier. They allow customers to purchase on credit in arbitrary time, location and amount, without carrying the cash.Provide a convenient payment method for purchases made on the internet, over the telephone, through ATMs, etc.
- **Keep customer credit history**
  Having a good credit history is often important in detecting loyal customers. This history is valuable not only for credit cards, but also for other financial serviceslike loans, rental applications, or even some jobs. Lenders and issuers of creditmortgage companies, credit card companies, retail stores, and utility companies can review customer credit score and history to see how punctualand responsible customers are in paying back their debts.
- **Protection of Purchases**
  Credit cards may also offer customers, additional protection if the purchased merchandisebecomes lost, damaged, or stolen. Both the buyer's credit card statement and company can confirmthat the customer has bought if the original receipt is lost or stolen. In addition, some credit card companies provideinsurance forlarge purchases.

In spite of all mentioned advantages, the problem of fraud is a serious issue ine-banking services that threaten credit card transactions especially. Fraud is an intentional deceptionwith the purpose of obtaining financial gain or causing loss by implicit or explicit trick.Fraud is a public law violation in which the fraudster gains an unlawful advantage or causes unlawful damage. The estimation ofamount of damage made by fraud activities indicates that fraud costs a very considerable sum of money.Credit card fraud is increasing significantly with the development of modern technology resulting in the loss of billions of dollars worldwide each year.Statistics from the Internet Crime Complaint Center show that there has been a significant rising in reported fraud in last decade.
Financial losses caused due to online fraud only in US, was reported $3.4 billion in 2011.
Fraud detection involves identifying scarce fraud activities among numerous legitimate transactions as quickly as possible. Fraud detection methods are developing rapidlyin order to adapt with new incoming fraudulent strategies across the world. But, development of new fraud detection techniquesbecomes more difficult due to the severe limitation of the ideas exchange in fraud detection. On the other hand, fraud detection is essentially a rare event problem, which has been variously called outlier analysis, anomaly detection, exception mining, mining rare classes, mining imbalanced data etc. The number of fraudulent transactions is usually a very low fraction of the total transactions. Hence the task of detecting fraud transactions in an accurate and efficient manner is fairly difficult and challengeable.Therefore, development of efficient methods which can distinguish rarefraud activities from billions of legitimate transaction seems essential.
Although, credit card fraud detection has gained attention and extensive studyespecially in recent years and there are lots of surveys about this kind of fraud such as [1], [2], [3],neither classify all credit card fraud detection techniques with analysis of datasets and attributes. Therefore in this paper, we attempt to collect and integrate a complete set of researches of literature and analyze them from various aspects.
The main contributions of this work are highlighted as follows:
- To the best of our knowledge, the absence of complete and detailed credit card fraud detection survey is an important issue, which is addressed by analyzing the state of the art in credit card fraud detection.

- The state of the art fraud detection techniques are described and classified from different aspects of supervised/unsupervised and numerical/categorical data consistent.

- In credit card fraud research each author has used its own dataset. There isno standard dataset or benchmark to evaluate detection methods. We attemptto gather different datasetsinvestigated by researchers, categorize them into real and synthetized groups and extract the common attributes affects the quality of detection.

The rest of the paper is organized as follows: a general description of types of fraud is presented in Section2.Challengesof credit card fraud detection are identified in Section3. Section4describes the credit card fraud detection techniques, their advantages and disadvantages and classification of them. In section 5 the dataset used by researchers and corresponding evaluation criteria are explained; another classification upon data types is also drawn in this section. Finally open issues of credit card fraud detection are presented in Section6.

## 2. Credit card fraud

Illegal use of credit card or its information without the knowledge of the owner is referred to as credit card fraud.Different credit card fraud tricks belong mainly to two groups of application and behavioral fraud [3]. Application fraud takes place when, fraudsters apply new cards from bank or issuing companies using false or other's information. Multiple applications may be submitted by one user with one set of user details (called duplication fraud) or different user with identical details (called identity fraud).

Behavioral fraud, on the other hand,has four principal types: stolen/lost card, mail theft, counterfeit card and 'card holder not present' fraud.Stolen/lost card fraud occurs when fraudsters steala credit card or get access to a lost card. Mail theft fraud occurs when the fraudster get a credit card in mail or personal information from bank before reaching to actual cardholder[3]. In both counterfeit and 'card holder not present' frauds, credit card details are obtained without the knowledge of card holders. In the former, remote transactions can be conducted using card details through mail, phone, or the Internet. In the latter, counterfeit cards are made based on card information.

Based on statistical data stated in [1] in 2012, the high risk countries facing credit card fraud threat is illustrated in Fig.1. Ukraine has the most fraud rate with staggering 19%, which is closely followed by Indonesia at 18.3% fraud rate. After these two, Yugoslavia with the rate of17.8% is the most risky country. The next highest fraud rate belongs to Malaysia (5.9%), Turkey (9%) and finally United States. Other countries that are prune to credit card fraud with the rate below than 1% are not demonstrated in figure 1.

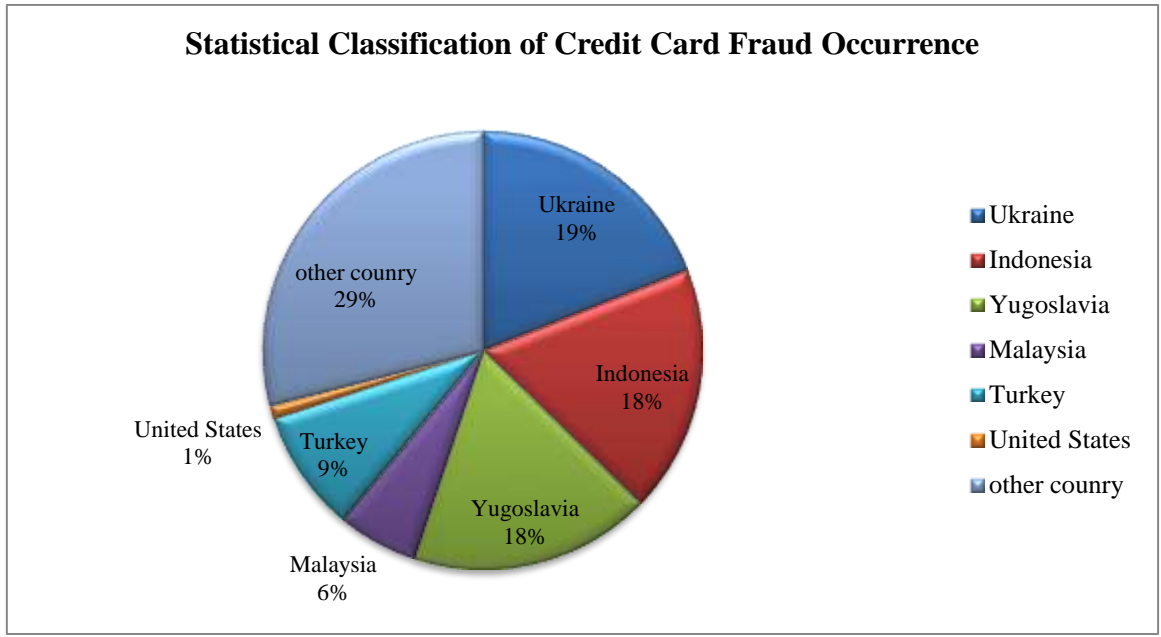

Fig1. High risk countries facing credit card fraud threat

## 3. Difficulties of Credit Card Fraud Detection

Fraud detection systems are prune to several difficulties and challenges enumerated bellow. An effective frauddetection technique should have abilities to address these difficulties in order to achievebest performance.

- **Imbalanced data**:The credit card fraud detection data has imbalanced nature. It means thatvery small percentages of all credit card transactions are fraudulent. This cause the detection of fraud transactions very difficult and imprecise.

- **Different misclassification importance:** in fraud detection task, different misclassification errors have different importance.Misclassification of a normal transaction asfraud is not as harmful as detecting a fraud transaction as normal. Because in the first case the mistake in classification will be identified in further investigations.

- **Overlapping data**: many transactions may be considered fraudulent, while actually they are normal (false positive) and reversely, a fraudulent transaction may also seem to be legitimate (false negative). Hence obtaining low rate of false positive and false negative is a key challenge of fraud detection systems[4, 5, and 6].

- **Lack of adaptability**: classification algorithms are usually faced with the problem of detecting new types of normal or fraudulent patterns. The supervised and unsupervised fraud detection systems are inefficient in detecting new patterns of normal and fraud behaviors, respectively.

- **Fraud detection cost**: The system should take into account both the cost of fraudulent behavior that is detected and the cost of preventing it. For example, no revenue is obtained by stopping a fraudulent transaction of a few dollars [5, 7].

- **Lack of standard metrics**: there is no standard evaluation criterion for assessing and comparing the results of fraud detection systems.

## 4. Credit Card Fraud Detection Techniques

The credit card fraud detection techniques are classified in two general categories: fraud analysis (misuse detection) and userbehavior analysis (anomaly detection).

The first group of techniques deals with supervised classification task in transaction level. In these methods, transactions are labeled as fraudulent or normal based on previous historical data. This dataset is then used to create classification models which can predict the state (normal or fraud) of new records. There are numerous model creation methods for a typical two class classification tasksuch as rule induction [1], decision trees [2] and neural networks [3].This approach is proven to reliably detect most fraud tricks which have been observed before [4], it also known as misuse detection.

The second approach deals with unsupervised methodologies which are based on account behavior. In this method a transaction is detected fraudulent if it is in contrast with user's normal behavior. This is because we don't expect fraudsters behave the same as the account owner or be aware of the behavior model of the owner [5].To this aim, we need to extract the legitimate user behavioral model (e.. user profile)for each account and then detect fraudulent activities according to it. Comparingnew behaviors with this model, different enough activities are distinguished as frauds. The profiles may contain the activity information of the account; such as merchant types, amount, location and time of transactions, [6].This method is also known as anomaly detection.

It is important to highlight the key differences between user behavior analysis and fraud analysis approaches.Thefraud analysis methodcan detect known fraud tricks, with a low false positiverate.These systems extract the signature and model of fraud tricks presented in oracle dataset and can then easily determine exactly which frauds, the system is currently experiencing. If the test data does not containanyfraud signatures, no alarm is raised. Thus, the false positive rate can be reduced extremely.However,sincelearning of a fraud analysis system (i.e. classifier) is based on limited and specific fraud records,It cannot detect novel frauds. As a result, the false negativesratemay be extremely high depending on how ingenious are the fraudsters.User behavioranalysis, on the other hand, greatly addresses the problem of detecting novel frauds. Thesemethods do not search for specific fraud patterns, but rather compare incoming activitieswiththe constructed model of legitimate user behavior. Any activity that is enough different from the model will be considered as a possible fraud. Though, user behavior analysis approaches are powerful in detectinginnovative frauds, they reallysuffer from high rates of false alarm. Moreover, if a fraud occurs during the training phase, this fraudulent behavior will be entered in baseline mode and is assumed to be normal in further analysis[7].In this section we will brieflyintroduce some current fraud detection techniques which are applied to credit card fraud detection tasks, also main advantage and disadvantage of each approachwill be discussed.

### 4.1 Artificial Neural Network
An artificial neural network (ANN) is a set of interconnected nodes designed to imitate the functioning of the human brain [9]. Each node has a weighted connection to several other nodes in adjacent layers. Individual nodes take the input received from connected nodes and use the weights together with a simple function to compute output values. Neural networks come in many shapes and architectures.The Neural network architecture, including the number of hidden layers, the number of nodeswithin a specific hidden layer and their connectivity, most be specified by user based on the complexity of the problem. ANNs can be configured by supervised, unsupervised or hybrid learning methods.

### 4.1.1 Supervised techniques

In supervised learning, samples of both fraudulent and non-fraudulent records, associated with their labels are used to create models. These techniques are often used in fraud analysis approach.One of the most popular supervised neural networks is back propagation network (BPN). It minimizes the objective function using a multi-stage dynamic optimization methodthat is a generalization of the delta rule.The back propagation method is often useful for feed-forward network with no feedback. The BPN algorithm is usually time-consuming and parameters like the number of hidden neurons and learning rate of delta rules require extensive tuning and trainingto achieve the best performance [10]. In the domain of fraud detection, supervised neural networks like back-propagation are known as efficient tool that have numerous applications [11], [12], [13].

RaghavendraPatidar, *et al*. [14] used a dataset to train a three layers backpropagation neural network in combination with genetic algorithms (GA)[15]forcredit card fraud detection. In this work, genetic algorithms was responsible for making decision about the network architecture, dealing with the network topology, number of hidden layers and number of nodes in each layer.

Also, Aleskerov*et al.* [16] developed a neural network based data mining system for credit card fraud detection.The proposed system (CARDWATCH) had three layers autoassociativearchitectures.They used a set of synthetized data for training and testing the system. The reportedresultsshow very successfulfraud detection rates.

In [17], a P-RCE neural network was applied for credit card fraud detection.P-RCE is a type of radial-basis function networks [18, 19]that usually applied for pattern recognition tasks.Krenker*et al*. proposed a model for real time fraud detection based on bidirectional neural networks [20]. They used a large data set of cell phone transactions provided by a credit card company. It was claimed that the system outperforms the rule based algorithms in terms of false positive rate.

Again in [21] a parallel granular neural network (GNN) is proposed to speed up data mining and knowledge discoveryprocess for credit card fraud detection.GNNis a kind of fuzzy neural network based on knowledge discovery (FNNKD).The underlying dataset was extracted from SQL server database containing sample Visa Card transactions and then preprocessed for applying in fraud detection. They obtained less average training errors in the presence of larger training dataset.

### 4.1.2 Unsupervised techniques

The unsupervised techniques do not need the previous knowledge of fraudulent and normalrecords.These methodsraise alarmfor those transactions that are most dissimilar from the normalones.These techniques are often used in user behavior approach.ANNs can produce acceptable result for enough large transaction dataset. They need a long training dataset. Self-organizing map (SOM) is one of the most popular unsupervised neural networks learning which was introduced by [22]. SOM provides a clustering method, which is appropriate for constructing and analyzing customer profiles,in credit card fraud detection, as suggested in [23]. SOM operates in two phase: training and mapping. In the former phase, the map is built and weights of the neurons are updated iteratively, based on input samples [24], in latter, test data is classified automatically into normal and fraudulent classes through the procedure of mapping. As stated in [25], after training the SOM, new unseen transactions are compared to normal and fraud clusters, if it is similar to all normal records, it is classified as normal. New fraud transactions are also detected similarly.

One of the advantages of using unsupervised neural networks over similar techniques is that these methods can learn from data stream. The more data passed to a SOM model, the more adaptation and improvement on result is obtained. More specifically, the SOM adapts its model as time passes. Therefore it can be used and updated online in banks or other financial corporations. As a result, the fraudulent use of a card can be detected fast and effectively. However, neural networks has some drawbacks and difficulties which are mainly related to specifying suitable architecture in one hand and excessive training required for reaching to best performance in other hand.

### 4. 1.3 Hybrid supervised and unsupervisedtechniques
In addition to supervised and unsupervised learning models of neural networks, some researchers have applied hybrid models. John ZhongLeiet.Al.[26] proposed hybrid supervised (SICLN) and unsupervised (ICLN)learning networkfor credit card fraud detection. They improved the reward only rule of SICLNmodel to ICLN in order to update weights according to both reward and penalty. This improvement appeared in terms of increasing stability and reducing the training time.Moreover, the number of final clusters of the ICLN is independent from the number of initial network neurons. As a result the inoperable neurons can be omitted from the clusters by applying the penalty rule.The results indicated that both the ICLN and the SICLN havehigh performance, but the SICLN outperforms well-known unsupervised clustering algorithms.

### 4.2 Artificial Immune System (AIS)
The natural immune system is a highly complex system, comprised of an intricate network of specialized tissues, organs, cells and chemical molecules. These elements are interrelated and act in a highly co- ordinate and specific manner when they recognize, remember disease causing foreign cells and eliminate them. Any element that could be recognized by the immune system is named an antigen. The immune system's detectors are the antibodies that are capable to recognition and destruction harmful and risky antigens [27].

The immune system consists of the two main response of immune and defense: innate immune response and acquired immune response. The body's first response for defense is made of the outer, unbroken skin and the 'mucus membranes' lining internal channels, such as the respiratory and digestive tracts. If the harmful cells could pass through innate immune defense the acquired immunity will defense. In fact, adaptive immune response performs based on antigen-specific recognition of almost unlimited types of infectious substances, even if previously unseen or mutated. It is worth mentioning that the acquired immune response is capable of "remembering" every infection, so that a second exposure to the same pathogen is dealt with more efficiently.

There are two organs responsible for the generation and development of immune cells: the bone marrow and the thymus. The bone marrow is the site where all blood cells are generated and where some of them are developed. The thymus is the organ to which a class of immune cells named T-cells migrates and maturates [28]. There exist a great number of different immune cells, but lymphocytes (white blood cells), are the prevailing ones. Their main function is distinguishing self-cells, which are the human body cells, from non-self cells, the dangerous foreign cells (the pathogens). Lymphocytes are classified into two main types: B-cells and T-cells, both originated in the bone marrow. Those lymphocytes that develop within the bone marrow are named B-cells, and those that migrate to and develop within the thymus (the organ which is located behind the breastbone) are named T-cells.

Artificial Immune System (AIS) is a recent sub field based on the biological metaphor of the immune system [29]. The immune system can distinguish between self and non-self-cells, or more specific, between harmful cells (called as pathogens) and other cells. The ability to recognize differences in patterns and being all to detect and eliminate infections precisely has attracted the engineer's intention in all fields.

Researchers have used the concepts of immunology in order to develop a set of algorithms, such as negative selection algorithm [30], immune networks algorithm [31], clonal selection algorithm [32], and the dendritic cells algorithm [33].

### 4.2.1 Negative Selection:
Negative Selection Algorithm or NSA proposed by [34] is a change detection algorithm based on the T-Cells generation process of biological immune system. It is one of the earliest AIS algorithms applied in various real-world applications. Since it was first conceived, it has attracted many researchers and practitioners in AIS and has gone through some phenomenal evolution. NSA has two stages: generation and detection. In generation stage, the detectors are generated by some random process and censored by trying to match self samples. Those candidates that match (by affinity of higher than affinity threshold) are eliminated and the rest are kept as detectors. In detection stage, the collection of detectors (or detector set) is used in checking whether an incoming data instance is self or non-self. If it matches (by affinity of higher than affinity threshold) any detector, it is claimed as non-self or anomaly.

Brabazon*et.al* [35] proposed an AIS based model for online credit card fraud detection. Three AIS algorithms were implemented and their performance was standardized against a logistic regression model. Their three chosenalgorithms were the unmodified negative selectionAlgorithm, the modified negative selection algorithm andthe Clonal selection algorithm.They proposed the Distance Value Metric for calculating distance between records. This metric is based on the probability of data occurrence in the training set. Where the detection rate increased, but the number of false alarms and missed frauds remained.

### 4.2.2 Clonal selection:
Clonal selection theory is used by the immune system to explain the basic features of an immune response to an antigenic stimulus. The selection mechanism guarantees that only those clones (antibodies) with higher affinity for the encountered antigen will survive. On the basis of clonal selection principle, clonal selection algorithm was initially proposed in [36] and formally explained in [37]. The general algorithm was called CLONALG.

Gadi*et.al* in [36] applied the AIRS in fraud detection on credit card transactions. AIRS is a classification algorithm that is based on AIS whichapplies clonal selection to create detectors.AIRS generatesdetectors for all of the classes in the database and in detectionstage uses k Nearest Neighbor algorithm (also called K-NN)in order to classify eachrecord.They compared their method with other methods like the neural networks, Bayesian networks, and decision trees and claimed that, after improving the input parameters for all the methods, AIRS has show the best results of all, partly perhaps since the number of input parameters for AIRS is comparatively high. If we consider a particular training dataset, and set the parameters depending on the same database, the results indicate a tendency to improve. The experiment was carried out on Weka package.

Soltani*et.al*in [8] proposed AIRS on credit card fraud detection. Since AIRS has a long training time, authors have implemented the model in Cloud Computing environment to shorten this time. They had used MapReduce API which works based on Hadoop distributed file system, and runs the algorithm in parallel.

### 4.2.3 Immune Network:

The nature immune system is applied through the interactions between a huge numbers of different types of cells. Instead of using a central coordinator, the nature immune systems sustain the appropriate level of immune responses by maintaining the equilibrium status between antibody suppression and stimulation using idiotypes and paratopes antibodies [38], [39]. The first Artificial Immune Network (AIN) proposed by [40]. Neal M.*et.al* [41] introduced the AISFD, which adopted thetechniques developed by CBR (case based reasoning) communityand applied various methods borrowed from genetic algorithm andother techniques to clone the B cells (network nodes) for mortgage fraud detection.

### 4.2.4 Danger Theory:

The novel immune theory, named Danger Theory was proposed in 1994 [42]. It embarked from the concept that defined "self-non-self" in the traditional theories and emphasizes that the immune system does not respond to "non-self" but to danger. According to the theory a useful evolutionarily immune system should focus on those things that are foreign and dangerous, rather than on those that are simply foreign [43]. Danger is measured by damage inflicted to cells indicated by distress signals emitted when cells go through an unnatural death (necrosis).

Dendritic cells (DCs), part of the innate immune system, interact with antigens derived from the host tissue; therefore, the algorithm inspired by Danger Theory is named Dendritic cell algorithm. Dendritic cells control the state of adaptive immune system cells by emitting the following signals:
- PAMP (pathogen associated molecular pattern)
- Danger
- Safe
- Inflammation

PAMP is released from tissue cells following sudden necrotic cell death; actually, the presence of PAMP usually indicates an anomalous situation

The presence of Danger signals may or may not indicate an anomalous situation; however the probability of an anomaly is higher than the same, under normal circumstances

Safe signal act as an indicator of healthy tissue

Inflammation signal is classed as the molecules of an inflammatory response to tissue injury. In fact, the presence of this signal amplifies the above three signals.

DCs exist in a number of different states of maturity, depending on the type of environmental signal present in the surrounding fluid. They can exist in immature, semi-mature or mature forms. Initially, when a DC enters the tissue, it exists in an immature state. DCs which have the ability to present both the antigen and active T-cells are mature. For an immature DC to become mature it should be exposed to PAMP and danger signals predominantly. The immature DCs exposed to safe signals predominantly are termed "semi-mature"; they produce semi-mature DCs output signaling molecule, which has the ability to de-activate the T-cells. Exposure to PAMP, danger and safe signals lead to an increase in co-stimulatory molecules production, which in turn ends up in removal from the tissue and its migration to local lymph nodes.

### 4.2.5 Hybrid AIS or methods

Some researchers applied different algorithms (i.e. vaccination algorithm, CART and so on) by AIS algorithm which are presented below:

Wong [44] presents the AISCCFD prototype proposed to measure and manage the memory population and mutate detectors in real time. In their work both the two algorithms the vaccination and negative selection were combined. The results were tested for different fraud types. The proposed method demonstrated higher detection rates when vaccination algorithm was applied, but it failed to detect some types of fraud precisely.

Huang *et.al* [45] presented a novel hybrid Artificial Immune inspired model for fraud detection by combining triplealgorithms: CSPRA, the dendritic cell algorithm (DCA), and CART. Though their proposed method had high detection rate and low false alarm, their approach was focused on logging data and limited to VoD (video on demand) systems and not credit card transactions.

Ayara*et.al* [46] applied AIS to predict failures of ATM[1]. Their approach is enriched by adding a generation of new antibodies from the antigens that correspond to the unpredicted failures.

### 4.3 Genetic Algorithm (GA)

Inspired from natural evolution, Genetic algorithms (GA), were originally introduced by John Holland [15]. GA searches for optimum solution with a population of candidate solutions that are traditionally represented in the form of binary strings called chromosomes.

The basic idea is that the stronger members of the population have more chance to survive and reproduce. The strength of a solution is its capability to solve the underlying problem which is indicated by fitness. New generation is selected in proportion to fitness among previous population and newly created offspring. Normally, new offspring will be produced by applying genetic operators such as mutation and crossing over on some fitter members of current generation (parents). As generations progress, the solution are evolved and the average fitness of population increases.This process is repeated until some stopping criteria, (i.e. often passing a pre-specified number of generations) is satisfied.

Genetic Programming (GP)[47] is an extension of genetic algorithms that represent each individual by a tree rather than a bit string. Due to hierarchy nature of the tree, GP can produce various types of model such as mathematical functions, logical and arithmetic expressions, computer programs, networks structures, etc.

Genetic algorithms have been used in data mining tasks mainly for feature selection. It is also widely used in combination with other algorithms for parameter tuning and optimization. Due to availability of genetic algorithm code in different programming languages, it is a popular and strong algorithm in credit card fraud detection. However, GA is very expensive in consuming time and memory. Genetic programming has also various applications in data mining as classification tool.

---

[1] Automatic Teller Machine (ATM)

EkremDuman *et al*. developed a method for credit card fraud detection [48]. They defined a cost-sensitive objective function that assigned different cost to different misclassification errors (e.g. false positive, false negative). In this case, the goal of a classifier will be the minimization of overall cost instead of the number of misclassified transactions. This is due the fact that the correct classification of some transactions was more important than others. The utilized classifier in this work was a novel combination of the genetic algorithms and the scatter search. For evaluating the proposed method, it was applied to real data and showed promising result in comparison to literature. Analyzing the influence of the features in detecting fraud indicated that statistics of the popular and unpopular regions for a credit card holder is the most important feature. Authors excluded some type of features such as the MCC and country statistics from their study that resulted in less generality for typical fraud detection problem.

K.RamaKalyani*et al*. [49] presented a model of credit card fraud detection based on the principles of genetic algorithm. The goal of the approach was first developing a synthetizing algorithm for generating test data and then to detect fraudulent transaction with the proposed algorithm..

Bentley et al. [50]developed a genetic programming based fuzzy system to extract rules for classifying data tested on real home insurance claims and credit card transactions.

In [51], authors applied Genetic Programming to the prediction of the price in the stock market of Japan. The objective of the work was to make decision in stock market about the best stocks as well as the time and amount of stocks to sell or buy. The experimental results showed the superior performance of GP over neural networks.

### 4.4 Hidden Markov Model (HMM)
A Hidden Markov Model is a double embedded stochastic process which is applied to model much more complicated stochastic processes as compared to a traditional Markov model. The underlying system is assumed to be a Markov process with unobserved states. In simpler Markov models like Markov chains, states are definite transition probabilities are only unknown parameters. In contrast, the states of a HMM are hidden, but state dependent outputs are visible.

In credit card fraud detection a HMM is trained for modeling the normal behavior encoded in user profiles [52]. According to this model, a new incoming transaction will be classified to fraud if it is not accepted by model with sufficiently high probability.Each user profile contains a set of information about last 10 transactions of that user liketime; category and amount of for each transaction [52, 53, and 54].HMM produces high false positive rate [55].V. Bhusari et al. [56] utialized HMM for detecting credit card frauds with low false alarm. The proposed system was also scalable for processing huge number of transactions.

HMM can also be embedded in online fraud detection systems which receive transaction details and verify whether it is normal or fraudulent.If the system confirms the transaction to be malicious, an alarm is raised and related bank rejects that transaction. The responding cardholder may then be informed about possible card misuse.

### 4.5 Support Vector Machine (SVM)
Support vector machine (SVM)[57] is a supervised learning model with associated learning algorithms that can analyze and recognize patterns for classification and regression tasks[48]. SVM is a binary classifier. The basic idea of SVM was to find an optimal hyper-plane which can separate instances of two given classes, linearly. This hyper-plane was assumed to be located in the gap between some marginal instances called support vectors. Introducing the kernel functions, the idea was extended for linearly inseparable data. A kernel function represents the dot product of projections of two data points in a high dimensional space. It is a transform that disperses data by mapping from the input space to a new space (feature space) in which the instances are more likely to be linearly separable. Kernels, such as radial basis function (RBF), can be used to learn complex input spaces. In classification tasks, given a set of training instances, marked with the label of the associated class, the SVM training algorithm find a hyper-plane that can assign new incoming instances into one of two classes. The class prediction of each new data point is based on which side of the hyper-plane it falls on feature space.

SVM has been successfully applied to a broad range of applications such as [58] [59] [60]. In credit card fraud detection, Ghosh and Reilly [61] developed a model using SVMs and admired neural networks. In this research a three layer feed-forward RBF neural network applied for detecting fraudulent credit card transactions through only two passes required to churn out a fraud score in every two hours.

Tung-shou Chen *et al*. [62] proposed a binary support vector system (BSVS), in which support vectors were selected by means of the genetic algorithms (GA). In proposed model self-organizing map (SOM) was first applied to obtain a high true negative rate and BSVS was then used to better train the data according their distribution.

In [63], a classification model based on decision trees and support vector machines (SVM) was constructed respectively for detecting credit card fraud. The first comparative study among SVM and decision tree methods in credit card fraud detection with a real data set was performed in this paper. The results revealed that the decision tree classifiers such as CART outperform SVM in solving the problem under investigation.

Rongchang Chen *et al*. [64] suggested a novel questionnaire-responder transaction (QRT) approach with SVM for credit card fraud detection. The objective of this research was the usage of SVM as well as other approaches such as Over-sampling and majority voting for investigating the prediction accuracy of their method in fraud detection. The experimental results indicated that the QRT approach has high degree of efficiency in terms of prediction accuracy.

Qibei Lu *et al*. [65] established a credit card fraud detection model based on Class Weighted SVM. Employing Principal Component Analysis (PCA), they initially reduced data dimension to less synthetic composite features due to the high dimensionality of data. Then according to imbalance characteristics of data, an improved Imbalance Class Weighted SVM (ICW-SVM) was proposed.

### 4.6 Bayesian Network

A Bayesian network is a graphical model that represents conditional dependencies among random variables. The underlying graphical model is in the form of directed acyclic graph. Bayesian networks are usefulfor finding unknown probabilities given known probabilitiesin the presence of uncertainty [66]. Bayesian networks can play an important and effective role in modeling situations where some basic information is already known but incoming data is uncertain or partially unavailable [67], [68], [69].The goal of using Bayes rules is often the prediction of the class label associated to a given vector of features or attributes [70].Bayesian networks have been successfully applied to various fields of interest for instance churn prevention[71] in business,pattern recognition in vision[72], generation of diagnostic in medicine[73]and fault diagnosis [74] as well as forecasting [75] in power systems.Besides, these networks have also been used to detect anomaly and frauds in credit card transactions or telecommunication networks [76, 77, and 5].

In [70], two approaches are suggested for credit card fraud detection using Bayesian network.In the first, the fraudulent user behavior and in the second the legitimate (normal) user behavior are modeled by Bayesian network. The fraudulent behavior net is constructed fromexpert knowledge, while the legitimate net is set up in respect to available data from non fraudulent users. During operation, legitimatenet is adapted to a specific user based on emerging data.Classification of new transactions were simplyconducted by inserting it to both networksand then specify the type of behavior (legitimate/fraud) according to correspondingprobabilities. Applying Bayes rule, gives the probability of fraud for new transactions [78]. Again, Ezawa and Nortondeveloped a four-stage Bayesian network [79]. They claimed that lots of popular methods such as regression, K-nearest neighbor and neural networks takes too long time to be applicable in their data.

### 4.7 Fuzzy Logic Based System
Fuzzy logic based system is the system based on fuzzy rules. Fuzzy logic systems address the uncertainty of the input and output variables by defining fuzzy sets and numbers in order to express values in the form of linguistic variables (e.g. small, medium and large). Two important types of these systems are fuzzy neural network and fuzzy Darwinian system.

#### 4.7.1 Fuzzy Neural Network (FNN)
The aim of applying Fuzzy Neural Network (FNN) is to learn fromgreatnumber of uncertain and imprecise records of information, which is very common in real world applications [80]. Fuzzy neural networks proposedin [81] to accelerate rule induction for fraud detection in customer specific credit cards. In this research authors applied GNN (Granular Neural Network) method which implements fuzzy neural network based on knowledge discovery (FNNKD), for accelerating thetraining network and detecting fraudster in parallel.

#### 4.7.2 Fuzzy Darwinian System
Fuzzy Darwinian Detection [82] is a kind of Evolutionary-Fuzzy system that uses genetic programming in order to evolve fuzzy rules. Extracting the rules, the systemcan classify the transactions into fraudulent and normal. This system was composed of genetic programming (GP) unit in combination with fuzzy expert system. Results indicated that the proposed system has very high accuracy and low false positive rate in comparison with other techniques, but it is extremely expensive [83].

### 4.8 Expert Systems
Rules can be generated from information which are obtained from a human expert and stored in a rule-based system as IF-THEN rules. Knowledge base system or an expert system is the information which is stored in Knowledge base. The rules in the expert system appliedin order to perform operations on a data to inference to reach appropriate conclusion. Powerful and flexible solutions for many application problemsprovides by expert system. Financial analysis and fraud detection are one of the general areas which it can be apply. By applying expert system suspicious activity or transaction can be detected from deviations from "normal' spending patterns [84].
In [85] authors presented a model to detect credit card frauds in various payment channels. In their model fuzzy expert system gives the abnormal degree which determines how the new transaction is fraudulent in comparison with user behavioral. The fraud tendency weight is achieved by user behavioral analysis. So, this system is named FUZZGY. Also, another research [86] proposed expert system model to detect fraud for alert financial institutions.

## 4.9 Inductive logic programming (ILP)

ILP by using a set of positive and negative examples uses first order predicate logic to define a concept. This logic program is then used to classify new instances. Complex relationship among components or attributes can be easily expressed,in this approach of classification.The effectiveness of the system improves by domain knowledge which can be easily represented in an ILP system [87].Muggleton et al.[88] proposed the model applying labeled data in fraud detection which using relational learning approaches such as Inductive Logic Programming (ILP) and simple homophily-based classifiers on relational databases. Perlich, et al. [89] also propose novel target-dependent detection techniques for converting the relational learning problem into a conventional one.

## 4.10 Case-based reasoning (CBR)

Adapting solutions in order to solve previous problems and use them to solve new problems is the basic idea of CBR. In CBR, cases introduce as descriptions of past experience of human specialists and stored in a database which uses for later retrieval when the user encounters a new case with similar parameters. These cases can apply for classification purposes. A CBR system attempts to find a matching case when face with a new problem. In this method the model defined as the training data, and in test phase when a new case or instance is given to the model it looks in all the data to discover a subset of cases that are most similar to new case and uses them to predict the result.

Nearest neighbor matching algorithm usually applied with CBR, although there are several another algorithms which used with this approach such as [90].

Case-based reasoning is well documented both as the framework for hybrid frauddetection systems and as an inference engine in [91].

Also, E.b. Reategui applied hybrid approaches of CBR and NN which divides the task of fraud detection into two separate components and found that this multiple approach was more effective than either approach on its own [92]. In this model, CBR looks for best matches in the case base while an artificial neural net (ANN) learns patterns of use and misuse of credit cards. The case base included information such as transaction amounts, dates, place and type, theft date, and MCC (merchant category code). The hybrid CBR and ANN system reported a classification accuracy of 89% on a case base of 1606 cases.

Table1. Advantages and disadvantages of fraud detection methods

| Techniques | | Advantages | Disadvantages |
|---|---|---|---|
| **Artificial Neural Network (ANN)** | | Ability to learn from the past/lack of need to be reprogrammed/ Ability to extract rules and predict future activities based on the current situation/ High accuracy/ Portability/ high speed in detection/ the ability to generate code to be used in real-time systems/ the easiness to be built and operated/ Effectiveness in dealing with noisy data, in predicting patterns, in solving complex problems, and in processing new instances/Adaptability /Maintainability /knowledge discovery and data miming | Difficulty to confirm the structure/high processing time for large neural networks and excessive training/ poor explanation capability/ difficult to setup and operate/high expense/ non numerical data need to be converted and normalized/Sensitivity to data format. |
| **Artificial Immune System (AIS)** | | High capability in pattern recognition/powerful in Learning and memory/Self-organization/ easy in integration with other systems/dynamically changing coverage/ self Identity/ multilayered/ has diversity/ noise tolerance/ fault tolerance/ predator-prey dynamics/ Inexpensive / no need to training phase in DCA. | Need high training time in NSA/ poor in handle missing data in ClonalG and NSA |
| **Genetic Algorithm** | | Works well with noisy data/easy to integrate with other systems/ usually combined into other techniques to increase the performance of those techniques and optimize their parameters/ easy in build and operate/Inexpensive/fast in detection/ Adaptability/Maintainability/knowledge discovery and data miming | Requires extensive tool knowledge to set up and operate and difficult to understand. |
| **Hidden Markov Model (HMM)** | | Fast in detection | Highly expensive/ low accuracy/not scalable to large size data sets |
| **Support Vector Machines (SVM)** | | SVMs deliver a unique solution, since the optimality problem is convex/by choosing an appropriate generalization grade, SVMs can be robust, even when the training sample has some bias. | Poor in process largedataset/expensive/has low speed of detection/ medium accuracy/lack of transparency of results |
| **Bayesian Network** | | High processing and detection speed/high accuracy | Excessive training need/ expensive |
| **Fuzzy Logic Based System** | **Fuzzy Neural Network** | Very fast in detection/good accuracy | Expensive |
| | **Fuzzy Darwinian System** | Very high accuracy/ Maintainability | Has very low speed in detection/ High expensive |
| **Expert System** | | Easy to modify the KB/ easy to develop and build the system/ easy to manage complexity or missing information/high degree of accuracy/ explanation facilities/good performance/Rules from other techniques such as NN and DT can be extracted, modified, and stored in the KB. | Poor in handling missing information or unexpected data values/poor in process different data types /knowledge representation languages do not approach human flexibility/ poor in build and operate/ poor in integration |
| **Inductive logic programming (ILP)** | | Powerful in process different data types/ powerful modeling language that can model complex relationships/powerful in handle missing data | Has low predictive accuracy/extremely sensitive to noise/ their performance deteriorates rapidly in the presence of spurious data. |
| **Case based reasoning (CBR)** | | Useful in domain that has a large number of examples/ has the ability to work with incomplete or noisy data/effective/ flexible/ easy to update and maintain/ can be used in a hybrid approach. | May suffer from the problem of incomplete or noisy data. |
| **Decision tree (DT)** | | High flexibility/good haleness/ explainable/easy to implement/easy to display and to understand | Requirements to check each condition one by one. In fraud detection condition is transaction. |

In order to best comprise in fraud detection techniques we have been summarizes the advantages and disadvantages of the mentioned techniques, which demonstrated in Table1. It is important to make a point that primary version of this Table presented in [87].Finally,Fig. 2shows a complete classification of fraud detection techniques.

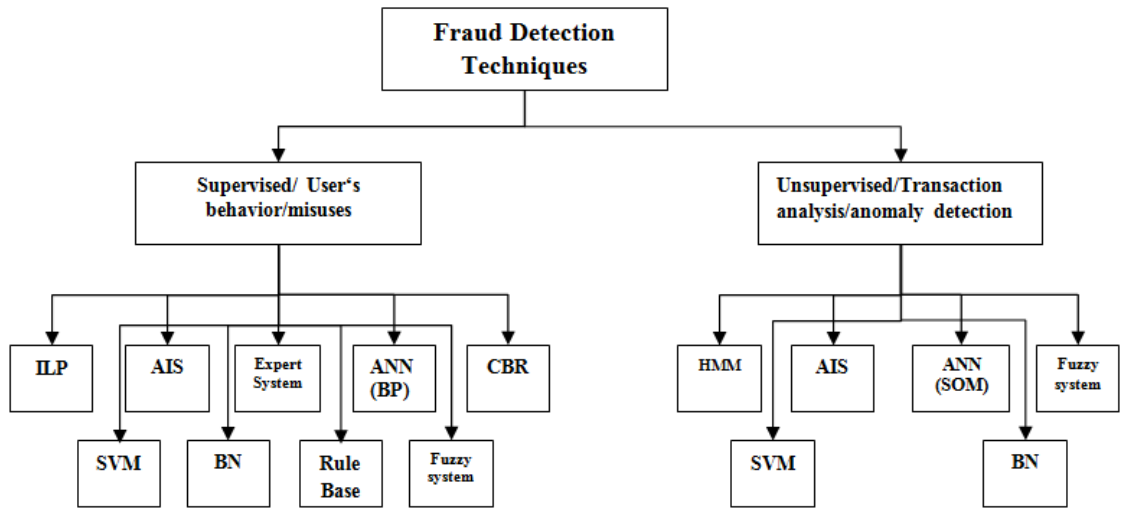

Fig. 2 A complete classification of credit card fraud detection techniques

**5-Data set and evaluation**

The mentioned methods in any field definitely need a creditable data set to test upon it, and examine efficiency in compare to other's related work. The lack of publicly available database has been a limiting factor for the publications on financial fraud detection [36], particularly credit card transactions. On the other hand, credit card is inherently private so, creating a proper data set for this purpose is very difficult and there are no standard techniques to do this.

Also, there is no universal corpus for credit card. Altogether, some works build their own data set for evaluation [48], [93]. However others use data sets which are gains by certain banks or financial institutions in a specific time window [36], [32], [94], [59], [95], and [96].

Table2 will be present some credit card fraud detection research which had used real or synthetically generated dataset.

Table2. Details of data set used by researchers

| Collection form | Amount | Used in | Methods |
|---|---|---|---|
| Large Brazilian bank, with registers within time window between Jul/14/2004 through Sep/12/2004. (Real Data set) | 41647 transactions/ 3.14% fraudulent transactions | Manoel Fernando, Alonso Gadi et al. (2008) | AIS |
| Financial institute in Ireland (WebBiz) (Real Data set) | 4 million transactions from 462279 unique customers/ 5417 fraudulent transactions | Anthony Brabazon et al. (2010) | AIS |
| Hong kong bank, with registers within time window between January 2006 to January 2007 (13 month) (Real Data set) | 50 million credit card transactions on about one million (1,167,757 credit cards) credit cards from a single country | C. Paasch (2007) / Siddhartha Bhattacharyya et al. (2010) | ANN tuned by genetic algorithm/ Data mining techniques |
| Chase Bank and First Union Bank (Real Data set) | Each bank supplied 500,000 records spanning one year with/ 20% fraud and 80% non fraud distribution for Chase Bank/ 15% versus 85% for First Union Bank | Philip K. Chan (1999) | Data mining techniques |
| Major US bank (Real Data set) | 6000 credit card data with 64 predictor variables plus 1 class variable, 84% of the data are normal accounts and 16% are fraudulent accounts | G. Kou, et al. (2005) | Multiple criteria linear programming |
| Large Australian bank (Real Data set) | 640361 total transactions, with 21746 credit cards | Nicholas Wong et al. (2012) | AIS |
| Vesta Corporation (Vesta corporation is an innovator and worldwide leader in virtual commerce with headquarter in Portland, Oregon, USA) (Real Data set) | 206,541 transactions, 204,078 transaction are normal and 2463 are fraudulent | John Zhong Lei (2012) | ANN |
| Mellon Bank (Real Data set) | 1,100,000 transactions/ authorized in two month period | SushmitoGhosh(1994) | ANN |
| Synthetically generated data | 320000000 transactions/ 1050 credit card/ 42 features | M. Hamdiozcelik et al. (2010) | GA |
| Synthetically generated data | 1000000 transactions/ 20 features | K.RamaKalyani et al. (2012) | GA |
| Synthetically generated data | 10000 transactions | Tao guo et al. (2008) | Data mining techniques |
| Synthetically generated data | The data are extracted into a flat file from SQL server database containing sample Visa Card transactions and then preprocessed. | MubeenaSyeda et al. (2002) | ANN |

Primary attributes are attributes of credit card transactions which are available in the most datasets. We present these mentioned attributes in Table3.

Table3. The common attributes in most datasets.

| Row | Attribute Name | Description |
|---|---|---|
| 1 | Posting date | Date when transaction was posted to the accounts |
| 2 | Merchant Category Code (MCC) | a code devote to each goods |
| 3 | Transaction Date and Time | Date and Time which the transaction was actually performed |
| 4 | Transaction status | status of transaction success/fail |
| 5 | Transaction Place | place of transaction (usually determine by IP Address) |
| 6 | Money Amount | Amount of Money |
| 7 | Transaction Type | Type of transaction payment/deposit/transfer and etc |
| 8 | customer identification | The identification code which advocate to each customer |
| 9 | Scheme | The type of credit card used, e.g. MasterCard, Visa, etc. |

The volume of fraud in every dataset is different. This might be because of the different security protocols used by different organizations and banks and so on. Whatever the reason is, this fact causes different fraud characteristics on each dataset, which affects the performance of the fraud detection system. Therefore considering the dataset's characteristics will help the system having more precise results.

A proper data set is a data set which covers various fraud and several attributes of customer profile or behavior. We believe that the contribution of attributes is a critical factor that should be considered. Also, a proper data set should be able to reflect the real world of credit card.

Credit card transaction datasets usually divided in to two types: numerical and categorical attributes. In statistics, categorical data is a statistical data type consisting of categorical variables, used for observed data whose value is one of a fixed number of nominal categories, or for data that has been converted into that form, for example as grouped data. However numeric data are numbers like age, cost, etc.

In fraud detection applications customer's gender and name are the typical numerical attribute, and categorical attributes are those like merchant category code, date of transaction, amount of transaction and etc. Some of these categorical variables can, depending on the dataset, have hundreds and thousands of categories.

Finally, Fig. 3 shows a complete classification in two groups: numerical and categorical attributes which is suitable for each algorithm.

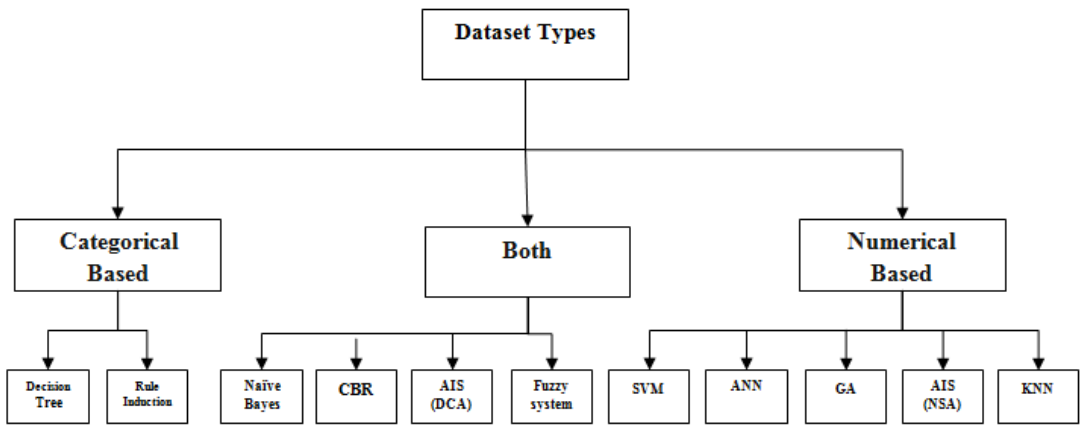

**Fig. 3** A complete classification of the dataset's attribute

- **Evaluation**

There are a variety of measures for various algorithms and these measures have been developed to evaluate very different things. So it should be criteria for evaluation of various proposed method. False Positive (FP), False Negative (FN), True Positive (TP), and True Negative (TN) and the relation between them are quantities which usually adopted by credit card fraud detection researcher to compare the accuracy of different approaches. The definitions of mentioned parameters are presented below:
- FP: the false positive rate indicates the portion of the non-fraudulent transactions wrongly being classified as fraudulent transactions.
- FN: the false negative rate indicates the portion of the fraudulent transactions wrongly being classified as normal transactions.
- TP: the true positive rate represents the portion of the fraudulent transactions correctly being classified as fraudulent transactions.
- TN: the true negative rate represents the portion of the normal transactions correctly being classified as normal transactions.

Table 4shows the details of the most common formulas which are used by researchers for evaluation of their proposed methods. As can be seen in this table some researchers had been used multiple formulas in order to evaluated their proposed model.

**Table4.** Evaluation criteria for credit card fraud detection

| Measure | Formula | Description | Used in |
|---|---|---|---|
| Accuracy (ACC)/Detection rate | TN + TP/TP + FP + FN + TN | Accuracy is the percentage of correctly classified instances. It is one the most widely used classification performance metrics | Nicholas Wong et al. (2012) [97], Manoel Fernando, et al. (2008)[36], soltani et al. [8],A. Brabazon et al. (2011) [35], Siddhartha et al. (2008) [59], P. Ravisankar et al. (2011) [99], AbhinavSrivastava et al. (2008) [52], John Zhong et al. (2012) [26], Qibei Lu et al. (2011) [65], AmlanKundu (2006) [98] |
| Precision/Hit rate | TP/TP + FP | Precision is the number of classified positive or fraudulent instances that actually are positive instances. | Manoel Fernando, et al. (2008)[36], Siddhartha et al. (2008) [50], John Zhong et al. (2012) [26], Qibei Lu et al. (2011) [65], AmlanKundu (2006) [98] |
| True positive rate/Sensitivity | TP/TP + FN | TP (true positive) is the number of correctly classified positive or abnormal instances. TP rate measures how well a classifier can recognize abnormal records. It is also called sensitivity measure. In the case of credit card fraud detection, abnormal instances are fraudulent transactions. | Maes S. et al. (2002) [5], Siddhartha et al. (2008) [59], Tao guo et al. (2008) [93], P. Ravisankar et al. (2011) [99], AbhinavSrivastava et al. (2008) [52], John Zhong et al. (2012) [26], Qibei Lu et al. (2011) [65], AmlanKundu (2006) [98] |
| True negative rate /Specificity | TN/TN + FP | TN (true negative) is the number of correctly classified negative or normal instances. TN rate measures how well a classifier can recognize normal records. It is also called specificity measure. | Siddhartha et al. (2008) [59], Philip K. Chan (1999) [95], Tao guo et al. (2008) [93], P. Ravisankar et al. (2011) [99], John Zhong et al. (2012) [26], Maes S. et al. (2002) [5], Qibei Lu et al. (2011) [65], AmlanKundu (2006) [98] |
| False positive rate (FPR) | FP/FP+TN | Ratio of credit card fraud detected incorrectly | Nicholas Wong et al. (2012) [97], soltani et al. [8], Maes S. et al. (2002) [5], Philip K. Chan (1999) [95], AbhinavSrivastava et al. (2008) [52], John Zhong et al. (2012) [26], Qibei Lu et al. (2011) [65], AmlanKundu (2006) [98] |
| ROC | True positive rate plotted against false positive rate | Relative Operating Characteristic curve, a comparison of TPR and FPR as the criterion changes | Manoel Fernando, et al. (2008)[36], Maes S. et al. (2002) [5], Tao guo et al. (2008) [93],John Zhong et al. (2012) [26], Qibei Lu et al. (2011) [65],AmlanKundu (2006) [98] |
| Cost | Cost = 100 * FN + 10 * (FP +TP) | | Manoel Fernando, et al. ,(2008)[36], soltani et al. [8], Philip K. Chan (1999) [95], Qibei Lu et al. (2011) [65] |
| F1-measure | 2 × (Precision ×Recall)/(Precision +Recall) | Weighted average of the precision and recall | Siddhartha et al. (2008) [59] |

The aim of all algorithms and techniques is to minimize FP and FN rate and maximize TP and TN rate and with a good detection rate at the same time.

## 6. Open issues

While credit card fraud detection has gained wide-scale attention in the literature, there are yet some issues (a number of significant open issues) that face researchers and have not been addressed beforeadequately. Wehope this overview focuses the direction of future research to provide more efficient and trustable fraud detection systems.These issues are as follow:

- **Nonexistence of standard and comprehensive credit card benchmark or dataset**

Credit card is inherently private property, so creating a proper benchmark for this purpose is very difficult. Incomplete datasets can cause fraud detection system to learn fraud tricks or normal behavior partially. On the other hand, lack of a standard dataset makes comparison of various techniques problematic or impossible. Many researchers used datasets that are only permitted to authors and cannot be published in order to privacy considerations.

- **Nonexistence of standard algorithm**

There is not any powerful algorithm known in credit card fraud literature that outperforms all others.Eachtechnique hasits own advantages and disadvantages as stated in previous sections. Combining these algorithms to support each other's benefits and cover their weaknesses would be of great interest.

- **Nonexistence of suitable metrics**

The limitation of good metrics in order to evaluate the results of fraud detection system is yet an open issue. Nonexistence of such metrics causes incapability of researchers and practitioners in comparing different approaches and determining priority of most efficient fraud detection systems.

- **Lack of adaptive credit card fraud detection systems**

Although lots of researches have been investigated credit card fraud detection field, there are none or limited adaptive techniques which can learn data stream of transactions as they are conducted. Such a system can update its internal model and mechanisms over a time without need to be relearned offline. Therefore, it can add novel frauds (or normal behaviors) immediately to model of learn fraud tricks and detect them afterward as soon as possible.